\documentclass[fleqn,twoside]{article}
\usepackage{espcrc2}

\usepackage{graphicx}
\usepackage[figuresright]{rotating}

\newcommand{\AmS}{{\protect\the\textfont2
  A\kern-.1667em\lower.5ex\hbox{M}\kern-.125emS}}

\newcommand{\Ref}[1]{Ref.~\cite{#1}}
\newcommand{\Refs}{Refs.}

\newcommand{\Eq}[1]{Eq.~(\ref{#1})}
\newcommand{\Fig}[1]{Fig.~\ref{#1}}

\title{Penta-Quark Anti-Decuplet in Anisotropic Lattice QCD}

\author{
  N.~Ishii\address[titech]{
    Dept. of Phys., H-27,
    Tokyo Institute of Technology,
    Meguro, Tokyo 152-8551, Japan}\thanks{e-mail: ishii@rarfaxp.riken.jp},
  T.~Doi\address{
    RIKEN BNL Research Center,
    Brookhaven National Laboratory,
    Upton, New York 11973, USA},
  H.~Iida\addressmark[titech],
  M.~Oka\addressmark,
  F.~Okiharu\address{
    Dept. of Phys.,
    Faculty of Science and Technology,
    Nihon Univ.,
    Chiyoda, Tokyo 101-8308, Japan},
  H.~Suganuma\addressmark[titech]
}
       
\begin{document}

\begin{abstract}
The penta-quark(5Q) $\Theta^+(1540)$ is studied in anisotropic lattice
QCD  with  renormalized anisotropy  $a_s/a_t=4$  for a  high-precision
measurement.
Both the positive and the negative parity 5Q baryons are studied using
a non-NK type interpolating field with $I=0$ and $J=1/2$.
After the  chiral extrapolation, the  lowest positive parity  state is
found  at  $m_{\Theta}\simeq 2.25$  GeV,  which  is  too heavy  to  be
identified with $\Theta^+(1540)$.
In the  negative parity channel, the  lowest energy state  is found at
$m_{\Theta}\simeq  1.75$ GeV.   Although  it is  rather  close to  the
empirical value, it is considered  to be an NK scattering state rather
than a localized resonance state.
\end{abstract}

\maketitle

LEPS group  at SPring-8 discovered the first  manifestly exotic hadron
$\Theta^+$ at  $1.54\pm 0.01$ GeV with  a width smaller  than $25$ MeV
\cite{nakano}.
The   experimental  discovery   \cite{nakano}  was   motivated   by  a
theoretical prediction \cite{diakonov}.
$\Theta^+$ is confirmed to have baryon number $B=1$, charge $Q=+1$ and
strangeness  $S=+1$, which  means that  it is  a baryon  containing at
least   one   $\bar{s}$.   Hence,   its   simplest  configuration   is
$uudd\bar{s}$, i.e., the penta-quark (5Q) state.

There   have  been   an   enormous  number   of  theoretical   studies
\cite{oka,zhu}  on $\Theta^+$ since  its discovery.   One of  the most
important  topics in 5Q  studies is  its parity.   Experimentally, the
parity    determination   of    $\Theta^+$   is    quite   challenging
\cite{hicks,thomas},  while opinions  are divided  in  the theoretical
side \cite{oka}.

There    are    several   lattice    QCD    studies   of    $\Theta^+$
\cite{fodor,sasaki,chiu,lee}, which have  not yet reached a consensus.
Except for  \Ref{chiu}, all  other calculations suggest  that negative
parity  states are  lighter than  positive parity  ones, and  that the
positive parity states are  quite massive.  \Ref{lee} has employed the
NK-type interpolating  field and  found no signal  on a  5Q resonance,
whereas   \Refs~\cite{fodor,sasaki}    have   employed   non-NK   type
interpolating fields and claimed the  existence of a 5Q resonance with
negative parity.
There  is  another  type of  lattice  QCD  studies  of the  static  5Q
potential  \cite{okiharu} aiming at  providing physical  insights into
the structure of penta-qurak baryons.

In   this  paper,  we   study  $\Theta^+$   for  both   parities  with
high-precision  data  generated  by  using  the  quenched  anisotropic
lattice  QCD.    We  employ  the  standard  Wilson   gauge  action  at
$\beta=5.75$  on the  $12^3\times  96$ lattice  with the  renormalized
anisotropy $a_s/a_t = 4$.   The anisotropic lattice technique is known
to     work    as     a    powerful     tool     for    high-precision
measurements \cite{klassen,matsufuru,nemoto,ishii}.
The  lattice spacing  is determined  from the  static  quark potential
adopting the Sommer parameter $r_0^{-1}= 395$ MeV leading to $a_s^{-1}
= 1.100(6)$ GeV ($a_s  \simeq 0.18$ fm) \cite{matsufuru}.  The lattice
size   $12^3\times   96$    amounts   to   $(2.15\mbox{fm})^3   \times
4.30\mbox{fm}$ in the physical unit.
For  the quark  part, we  employ the  $O(a)$-improved  Wilson (clover)
action  \cite{matsufuru} with  four  values of  hopping parameters  as
$\kappa=0.1210(0.0010)0.1240$, which correspond to $m_{\pi}/m_{\rho} =
0.81, 0.77, 0.72$ and $0.65$.
By   keeping  $\kappa_s=0.1240$   fixed   for  s   quark,  we   change
$\kappa=0.1210-0.1240$ for u and  d quarks for chiral extrapolation.
Anti-periodic  boundary  condition (BC)  is  imposed  on the  temporal
direction, whereas  periodic BC is  imposed on the  spatial directions
for quark fields.
To enhance the  low-lying spectra, we adopt a  smeared source with the
gaussian size $\rho\simeq 0.4$ fm.
We  use   504  gauge   configurations  to  construct   correlators  of
$\Theta^+$.
For detail, see \Ref{ishii-penta}.

We consider a non-NK type interpolating field for $\Theta^+$ as
\begin{equation}
  O
  \equiv
  \epsilon_{abc}
  \epsilon_{ade}
  \epsilon_{bfg}
  \left(u_d^T C\gamma_5 d_e\right)
  \left(u_f^T C d_g\right)
  \left(C\bar{s}_c^T\right),
  \label{sasaki-op}
\end{equation}
where  $a-g$  denote  color  indices, and  $C\equiv  \gamma_4\gamma_2$
denotes the charge conjugation matrix.
The quantum number of $O$ is spin $J=1/2$ and isospin $I=0$.
Under the spatial  reflection of the quark fields,  i.e., $q(t,\vec x)
\to \gamma_4  q(t,-\vec x)$, $O$  transforms exactly in the  same way,
i.e., $O(t,\vec x)  \to +\gamma_4 O(t,-\vec x)$, which  means that the
intrinsic parity of $O$ is positive.
Although  its intrinsic  parity is  positive, it  couples  to negative
parity states as well \cite{montvay}.

We  consider the  asymptotic behavior  of the  zero-momentum projected
correlator as
\begin{equation}
  G_{\alpha\beta}(t)
  \equiv
  \frac1{V}
  \sum_{\vec x}
  \left\langle
  O_{\alpha}(t,\vec x)
  \bar{O}_{\beta}(0,\vec 0)
  \right\rangle,
\end{equation}
where $V$ denotes  the spatial volume.  In the region of  $0 \ll t \ll
N_t$ with  $N_t$ being  the temporal lattice  size, the  correlator is
decomposed into two parts as
\begin{eqnarray}
  G(t)
  &\equiv&
  P_{+}
  \left( C_+ e^{-m_+ t} - C_- e^{-m_-(N_t - t)} \right)
  \label{spectral.rep}
  \\\nonumber
  &+&
  P_{-}
  \left( C_- e^{-m_- t} - C_+ e^{-m_+(N_t - t)} \right),
\end{eqnarray}
where  $m_{\pm}$  refer to  the  energies  of  lowest-lying states  in
positive and  negative parity channels,  respectively. $P_{\pm} \equiv
(1\pm \gamma_4)/2$ serve as projection matrices onto the ``upper'' and
``lower''  Dirac  subspaces,   respectively,  in  the  standard  Dirac
representation.   \Eq{spectral.rep} suggests  that, in  the  region of
$0\ll  t  \ll  N_t/2$,   the  backwardly  propagating  states  can  be
neglected.   Hence,  ``upper''  Dirac  subspace is  dominated  by  the
lowest-lying positive  parity state, whereas  ``lower'' Dirac subspace
is dominated  by the lowest-lying  negative parity state.   We utilize
this property for parity projection.

\begin{figure}[h]
\begin{center}
\hspace*{-1em}\includegraphics[height=0.5\textwidth,width=0.45\textwidth,angle=-90]{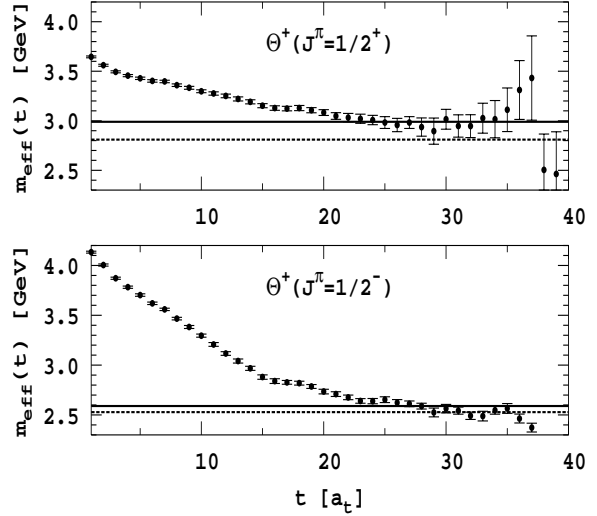}
\end{center}
\vspace{-0.95cm}
\caption{The  effective mass  plots  of positive  and negative  parity
$\Theta^+$ for a typical set  of the hopping parameters as $(\kappa_s,
\kappa) = (0.1240,0.1220)$.  The solid  lines denote the result of the
single-exponential fit performed in the  region, $25\le t \le 35$. The
dotted  lines  denote the  p-wave  (s-wave)  NK  threshold energy  for
positive  (negative) parity channels  on the  spatial lattice  size $L
\simeq 2.15$ fm.}
\label{fig.effmass}
\end{figure}
In \Fig{fig.effmass}, we show the effective mass plots for both parity
channels, which are  obtained from a correlator with  a smeared source
and a point sink, adopting a  typical set of the hopping parameters as
$(\kappa_s,\kappa) =  (0.1240, 0.1220)$.   For both channels,  we find
plateaus in the  region $25 \le t \le 35$. We  simply neglect the data
for $t >  35$, where backwardly propagating contributions  are seen to
become less  negligible.  The  single-exponential fit is  performed in
the  plateau region.   The results  are denoted  by solid  lines.  The
dotted lines  indicate the p-wave (s-wave) NK  thresholds for positive
(negative) parity channels on  the spatial lattice size $L\simeq 2.15$
fm. Note that due to the quantized spatial momentum in the finite box,
the p-wave  threshold is raised  as $E_{th} \simeq \sqrt{m_N^2  + \vec
p_{\rm min}^2} + \sqrt{m_K^2 + \vec p_{\rm min}^2}$ with $|\vec p_{\rm
min}| = 2\pi/L$.

\begin{figure}[h]
\begin{center}
\includegraphics[height=0.5\textwidth,angle=-90]{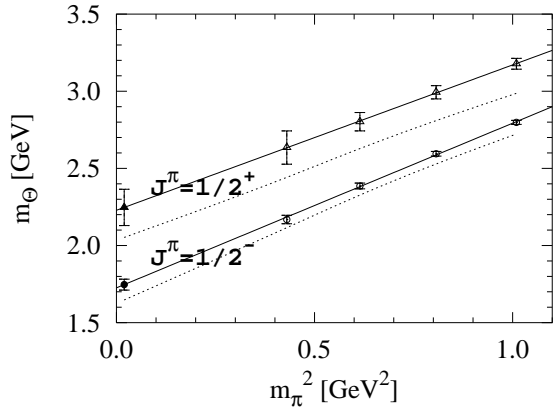}
\end{center}
\vspace{-0.95cm}
\caption{$m_{\Theta}$    for     both    parity    channels    against
$m_{\pi}^2$. The  triangles correspond  to the positive  parity, while
the circles correspond to the negative parity. The open symbols denote
direct lattice  data, whereas  the closed ones  the results  after the
chiral  extrapolation.  The  dotted  lines indicate  the NK  threshold
energies for p-wave (upper) and s-wave (lower) cases.}
\label{fig.chiral}
\end{figure}
In \Fig{fig.chiral},  the masses  of positive (triangle)  and negative
(circle) parity  $\Theta^+$ are plotted against  $m_{\pi}^2$. The open
symbols denote  direct lattice  data.  We find  that the  data behaves
linearly in $m_{\pi}^2$. Such a linear behavior against $m_{\pi}^2$ is
also    observed   for    ordinary   non-PS    mesons    and   baryons
\cite{matsufuru,nemoto}.  We extrapolate  the lattice data linearly to
the  physical quark  mass region.  The results  are denoted  by closed
symbols.  For  convenience, we show p-wave (upper)  and s-wave (lower)
NK threshold with dotted lines.

In  the positive  parity channel,  the chiral  extrapolation  leads to
$m_{\Theta}=2.25$ GeV,  which is much heavier  than the experimentally
observed $\Theta^+(1540)$.
In contrast, in the  negative parity channel, the chiral extrapolation
leads to $m_{\Theta}=1.75$ GeV, which is rather close to the empirical
value. However, from a recent progress using a new general method with
a  ``{\em hybrid  boundary  condition}'', it  turns  out to  be an  NK
scattering    state   rather   than    a   localized    5Q   resonance
\cite{ishii-penta}.

To summarize,  we have studied  the penta-quark $\Theta^+$  state with
the   anisotropic   lattice  QCD   at   quenched   level  to   provide
high-precision data.
After the chiral extrapolation,  we have obtained a massive $\Theta^+$
as $m_{\Theta}=2.25$ GeV in the positive parity channel.
We have therefore concluded that  this state cannot be identified with
the experimentally  observed $\Theta^+(1540)$.
On the  other hand, in the  negative parity channel,  we have obtained
$m_{\Theta}=1.75$ GeV,  which is rather close to  the empirical value.
However,  from a recent  progress using  a new  general method  with a
``{\em hybrid boundary  condition}'', we have concluded that  it is an
NK scattering state. For detail, see \Ref{ishii-penta}.

\vspace{0.3cm}
\hspace*{-1em}{\bf  Acknowledgements}\\
Lattice  QCD Monte  Carlo calculations  have been  done on  NEC-SX5 at
Osaka University.


\begin{thebibliography}{9}
\bibitem{nakano}
  LEPS Collaboration, T.~Nakano {\it et al.},
  Phys.~Rev.~Lett.~{\bf 91} (2003) 012002.
\bibitem{diakonov}
  D.~Diakonov, V.~Petrov and M.V.~Polyakov,
  Z. Phys. {\bf A359} (1997) 305.
\bibitem{oka}
  For a review article,
  M.~Oka,
  Prog. Theor. Phys. {\bf 111} (2004) 1, and references therein.
\bibitem{zhu}
  S.L. Zhu, hep-ph/0406204 and its references.
\bibitem{hicks}
  T.~Nakano and K.~Hicks,
  Mod. Phys. Lett. {\bf A19} (2004) 645.
\bibitem{thomas}
  A.W.~Thomas, K.~Hicks and A.~Hosaka,
  Prog. Theor. Phys. {\bf 111} (2004) 291.
\bibitem{fodor}
  F.~Csikor, Z.~Fodor, S.D.~Katz and T.G.~Kovacs,
  JHEP {\bf 0311} (2003) 070.
\bibitem{sasaki}
  S.~Sasaki,
  in this proceedings.
\bibitem{chiu}
  T.W.~Chiu and T.H.~Hsieh,
  hep-ph/0403020.
\bibitem{lee}
  N.~Mathur, F.X.~Lee, A.~Alexandru, C.~Bennhold,
  Y.~Chen, S.J.~Dong, T.~Draper, I.~Horv\'ath,
  K.F.~Liu, S.~Tamhankar and J.B.~Zang,
  hep-ph/0406196.
\bibitem{okiharu}
  H.~Suganuma, T.T.~Takahashi, F.~Okiharu and H.~Ichie,
  Proc. of QCD Down Under, Adelaide, March 2004,
  Nucl. Phys. {\bf B} (Proc. Suppl.) in press;
  F.~Okiharu, H.~Suganuma and T.T.~Takahashi,
  hep-lat/0407001.  
\bibitem{klassen}
  T.R.~Klassen,
  Nucl. Phys. {\bf B533} (1998) 557.
\bibitem{matsufuru}
  H.~Matsufuru, T.~Onogi and T.~Umeda,
  Phys. Rev. {\bf D64} (2001) 114503.
\bibitem{nemoto}
  Y.~Nemoto, N.~Nakajima, H.~Matsufuru and H.~Suganuma,
  Phys. Rev. {\bf D68} (2003) 094505.
\bibitem{ishii}
  N.~Ishii, H.~Suganuma and H.~Matsufuru,
  Phys. Rev. {\bf D66} (2002) 094506; Phys. Rev. {\bf D66} (2002) 014507.
\bibitem{ishii-penta}
  N.~Ishii, T.~Doi, H.~Iida, M.~Oka, F.~Okiharu and H.~Suganuma,
  hep-lat/0408030.
\bibitem{montvay}
  I.~Montvay and G.~M\"unster,
  ``{\it Quantum Fields on a Lattice}'',
  (Cambridge Univ. Press, Cambridge, England, 1994), p.~1.
\end{thebibliography}
\end{document}